\documentstyle[preprint,aps]{revtex}
\begin{document}
\title{\bf Universality in Sandpile Models}
\author{Asa Ben-Hur\footnote{E-mail: asa@ariel.fiz.huji.ac.il}
 and Ofer Biham\footnote{E-mail: biham@cc.huji.ac.il}}
\address{
Racah Institute of Physics,
The Hebrew University,
Jerusalem 91904,
Israel}

\maketitle

\begin{abstract}
A new classification of sandpile models into universality classes
 is presented.
On the basis of extensive numerical simulations, in which we
 measure an extended set of exponents,
the Manna two state model [S. S. Manna,  J. Phys. A 24,  L363 (1991)] is found to belong to a universality
 class of random neighbor models which is distinct from the
 universality class of the original model of Bak, Tang and
 Wiesenfeld [P. Bak,  C. Tang and K. Wiensenfeld, 
 Phys. Rev. Lett. 59,  381 (1987)].
Directed models are found to belong to a universality class which
 includes the directed model introduced and solved by Dhar
and Ramaswamy.

\end{abstract}


\newpage

The introduction of sandpile models as a paradigm of self-organized
 criticality by Bak, Tang and Wiesenfeld (BTW)~\cite{btw} stimulated 
numerous theoretical~\cite{dhar,dhar-dir} and numerical 
studies~\cite{manna,gras,christ-fogedby,christ-olami}.
In these models, which are defined on a lattice, grains are 
deposited randomly until the height at some site exceeds a 
threshold, and becomes unstable.  
``Sand'' is then distributed to the nearest neighbors.  
As a result of this relaxation process neighboring sites may become
 unstable, resulting in a cascade of relaxations called {\it an 
avalanche}.
It was observed that these models are self-driven into a critical 
state which is characterized by a set of exponents~\cite{btw}.
These include exponents that describe the distribution of 
quantities such as avalanche size and lifetime, and exponents which 
relate these properties of the dynamics.
Large scale simulations of the BTW model~\cite{manna} and some 
variants of it~\cite{twostate,cascades} were performed.
The BTW model and the Manna two-state model were concluded to 
belong to the same universality class~\cite{twostate}.
Christensen and Olami later introduced an extended set of 
exponents~\cite{christ-olami}.
They measured the values of these exponents for the BTW model, and
 gave theoretical predictions and heuristic arguments for the
 values of some of the exponents.
Continuous height models were also studied~\cite{zhang} and some
 aspects of universality were examined~\cite{diaz}.
A sandpile model with a preferred direction was introduced and 
solved by Dhar and Ramaswamy~\cite{dhar-dir}.

In this paper we present simulation results which suggest a new 
classification of sandpile models into universality classes.
The Manna two-state model is found to belong to a universality 
class of random relaxation models which is distinct from the
BTW 
universality class.
We first describe the different models, and define the properties 
of avalanches, with the exponents characterizing them.
The models are defined on a $d$ dimensional lattice of linear size 
$L$.
Each site is assigned a dynamic variable $E(\bf i)$ which 
represents some physical quantity such as energy, stress etc.
In a critical height model a configuration
 $\{E({\bf i})\}$ is called {\it stable} if
for all sites $E({\bf i}) < E_c$, where $E_c$ is a threshold value.
The evolution between stable configurations is by the following 
rules:

 (i) Adding energy.
Given an arbitrary stable configuration $\{E(\bf j)\}$ we select a 
site ${\bf i}$ at random and increase $E({\bf i})$ by some amount 
$\delta E$.
When an unstable configuration is reached rule (ii) is invoked.

(ii) The relaxation rule.
If the dynamical variable at site ${\bf i}$ exceeds the 
threshold $E_c$, relaxation takes place, whereby energy is 
distributed in the following way:

\begin{eqnarray}\label{def}
E({\bf i}) & \rightarrow & E({\bf i}) - \sum_{\bf e}\Delta 
E({\bf e})  \nonumber \\
\\				
E({\bf i}+{\bf e}) & \rightarrow & E({\bf i}+{\bf e})+
\Delta E({\bf e}),\nonumber 
\end{eqnarray}
where ${\bf e}$ are a set of (unit) vectors from the site ${\bf i}$
to some neighbors. 
As a result of the relaxation the dynamic variable in one or more of
 the neighbors may exceed the threshold.  
The relaxation rule is then applied until a stable 
configuration is reached.
The sequence of relaxations is an avalanche which propagates 
through the lattice.

The parameters $\delta E$ and $E_c$  are irrelevant to the scaling 
behavior~\cite{dhar,diaz}.
Thus the only factor determining the exponents is the
vector $\Delta E$, to be termed {\it relaxation vector}. 
For a square lattice with relaxation to nearest neighbors it is of 
the form $\Delta E=(E_N,E_E,E_S,E_W)$, where $E_N$ for example is 
the amount transferred to the northern nearest neighbor.
The original BTW model is given by the vector $(1,1,1,1)$.
The relaxation in the directed 
model of Dhar and Ramaswamy~\cite{dhar-dir} is 
specified by any vector with ones in two adjacent directions and 
zeroes in the two other directions, such as $(0,0,1,1)$.
In a random relaxation model a set of neighbors is randomly chosen 
for relaxation.
Such a model is specified by a set of relaxation vectors, each 
vector being assigned a probability for its application.
As an example, a possible realization of a two-state model makes 
use of the six relaxation vectors (1,1,0,0),(1,0,1,0),(1,0,0,1),
(0,1,1,0),(0,1,0,1) and (0,0,1,1), each one applied with a 
probability of $1/6$.
In Manna's two-state model~\cite{twostate} the variable is 
decreased to zero on relaxation, with sand distributed randomly 
among the nearest neighbors.
We define a current
\begin{equation} \label{current}
	{\bf J}[\Delta E]=\sum_{\bf e} \Delta E({\bf e}){\bf e},
\end{equation}
which is the net flow in a relaxation.
We also define
\begin{equation} \label{avg-current}
	{\bf J}= \sum_{\Delta E} {\bf J}[\Delta E]P(\Delta E),
\end{equation}
which is the current averaged over the ensemble of relaxation
vectors. 
Models can be classified according to the value of the current 
${\bf J}[\Delta E]$ and its average, {\bf J}.
A model is called {\it non-directed} if ${\bf J}[\Delta E]=0$ 
(the BTW model for example).
Random relaxation models such as the Manna two-state model, 
which satisfy
${\bf J}[\Delta E]\neq 0$ 
and 
${\bf J}=0$,
are called 
{\it non-directed on average}. 
Models with ${\bf J}\neq 0$ are called {\it directed}.  
In this paper we present evidence that this is a classification into 
universality classes.

Avalanches have various properties which can be measured in a 
simulation:
size, area, lifetime, linear size, and perimeter.
The size ($s$) of an avalanche is the total number of relaxation 
events that occured in the course of a single avalanche.
The area ($a$) is the number of sites in the lattice where 
relaxation occured.
Relaxation of all sites which exceed the threshold at a given time 
is considered a single time step.
The lifetime ($t$) of an avalanche is the number of such steps.
As for the linear size of an avalanche, there is no unique choice.
A possible choice is the maximal distance ($d$) between the origin 
of the avalanche
to sites of the avalanche cluster.
Another possibility is the radius of gyration ($r$) of the cluster 
of sites where relaxation occured.
A site belonging to the cluster of sites visited by an avalanche is
 defined to be a perimeter site if it has a nearest neighbor where 
no relaxation took place.
The perimeter ($p$) is the number of perimeter sites. 
Thus we have a set of variables $\{s,a,t,r,d,p\}$ which 
characterize an avalanche.
The avalanche variables have probability functions which are assumed
 to fall off with a power law defined by
$P(x) \sim x^{1-\tau_{x}}$,
where $x \in \{s,a,t,r,d,p\}$.
These variables also scale against each other in the form
\begin{equation}
	y \sim x^{\gamma_{yx}},
\end{equation}
for $x,y \in \{s,a,t,r,d,p\}$.
The exact definition of the $\gamma$'s is in terms of conditional 
expectations values: 
$E[y | x] \sim x^{\gamma_{yx}}$~\cite{christ-fogedby}.
The exponents are not independent.
Scaling relations are found 
in~\cite{christ-olami}.
We just note that 
\begin{eqnarray}\label{gamma}
\gamma_{yx}&=&\gamma_{xy}^{-1}, \nonumber \\
\\ 
\gamma_{zx}&=&\gamma_{zy}\gamma_{yx}. \nonumber
\end{eqnarray}
Avalanches are proven to be compact for BTW type 
models~\cite{christ-olami} but have a fractal boundary.
It is reasonable to assume that the fractal dimension $D_f$ of the 
boundary is given by the scaling of the perimeter (p) against the 
linear size of the avalanche.
It seems that for models which are non-directed the radius of 
gyration is the proper measure of size~\cite{diaz}.  
Therefore we identify $D_f$ with $\gamma_{pr}$.
For directed models the maximum distance from the origin to the 
perimeter is the proper measure of size, and $D_f$ is identified 
with $\gamma_{pd}$.
It is accepted that the dynamical exponent $z$ of non-directed 
models should be identified with $\gamma_{tr}$~\cite{diaz}.
In the case of directed models we identify the dynamical exponent 
with $\gamma_{td}$.

Having defined the models, we now describe the simulations.
We used open boundary conditions and system sizes up to $512^2$, 
 with 5 million grains dropped, in two dimensions; 
in three dimensions system sizes were
up to $112^3$, with 20 million grains dropped.
An algorithm due to Grassberger~\cite{gras} was used.
We ascertained the dynamics has reached the critical state by 
applying Dhar's ``burning algorithm''~\cite{dhar}, or by starting
 with a configuration belonging to the critical state.
Manna's and our own simulation results for the BTW model indicate 
that the distribution exponents are system size dependent,  
with a logarithmic convergence to the infinite system values.
The values of the $\gamma$'s on the other hand, seem to be almost 
independent of system size.
Moreover, we found that the relations that specify the $\gamma$'s
 hold during avalanches as well, and are not just
a scaling property of completed avalanches.
Thus the $\gamma$'s provide a robust characterization of the 
dynamical properties of a sandpile model,
and can be used for a reliable classification of sandpile models 
into universality classes.

Previous studies clearly show that directed and non-directed models
 belong to different universality classes
~\cite{dhar-dir,christ-olami,twostate}.
On the basis of Manna's simulation results it was concluded that 
the Manna two-state model and the BTW model are in the same 
universality class
~\cite{twostate}.
This conclusion is based on measurements of a limited set of 
exponents:  $\tau_s$,$\tau_t$ and $\gamma_{ts}$.
We measured the extended set of exponents introduced by 
Christensen and Olami, and the fractal dimension.
The $\gamma$'s we obtained in two dimensions are listed in 
Table~\ref{tablegamma}.
Our results are consistent with known analytical results and 
simulation data:
Dhar and Ramaswamy's analytical solution 
of a directed model~\cite{dhar-dir};
simulation results and scaling arguments given by Christensen and 
Olami~\cite{christ-olami};
simulation results of Manna~\cite{manna,twostate}.
A momentum-space analysis of a Langevin equation indicates that 
for the BTW model $z=(2+d)/3$~\cite{diaz}.  Our results for 
$\gamma_{rt}$ which is identified with $1/z$, confirm this scaling 
relation.
This agreement supports our observation that the $\gamma$'s are 
size independent, and indicates that we are in the right avalanche 
size regime for the observation of $\gamma_{rt}$.
On the basis of the difference in the $\gamma$'s for the BTW and 
two-state models we conclude that the two models are not in the 
same universality class
(Fig.~\ref{figcomparison}).

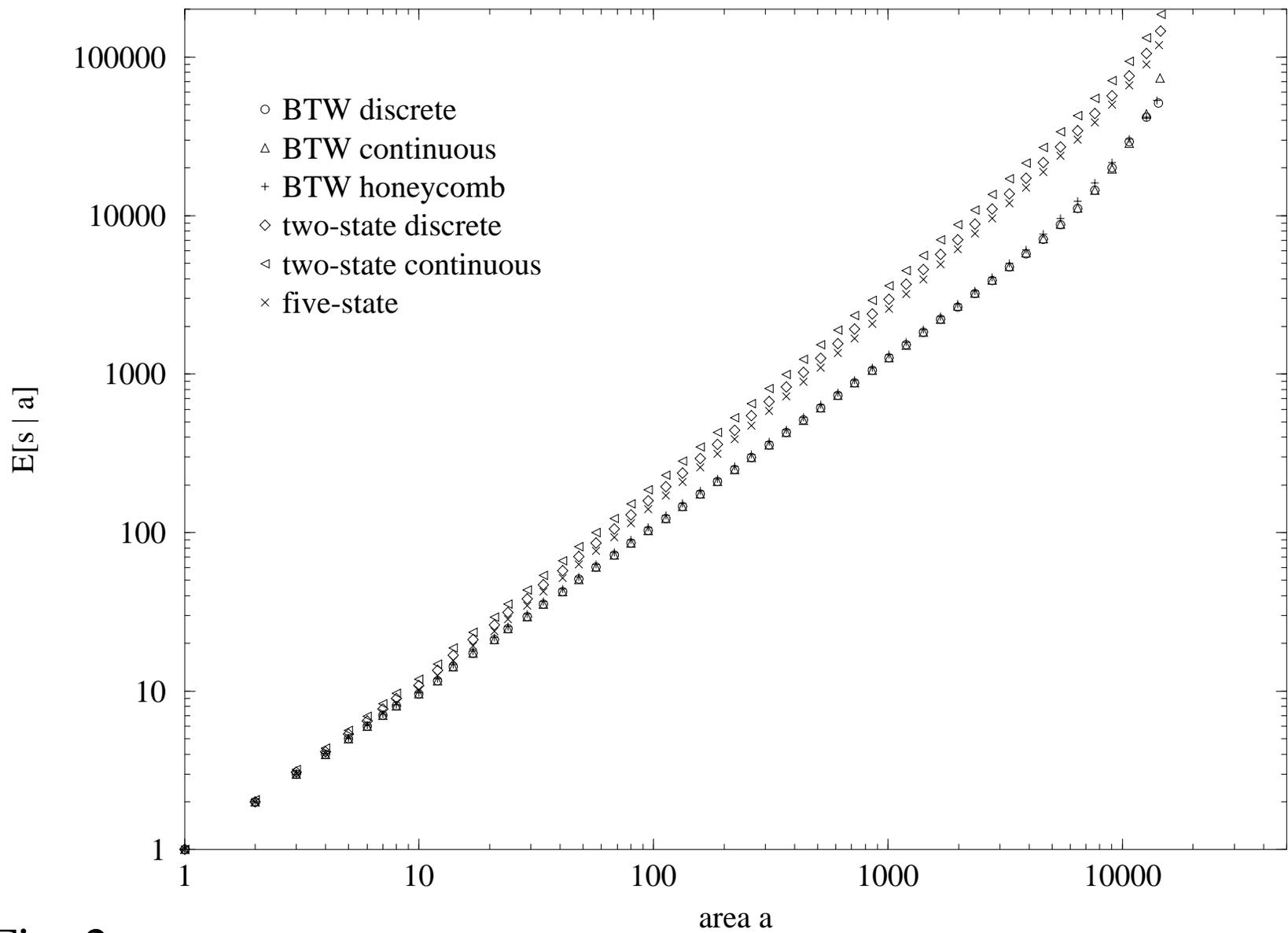
\begin{figure}
\caption{Simulation results showing the universality of models in
 the BTW (non-directed)
and two-state (non-directed on average) universality classes.
Graph shows $E[s | a]$ vs. $a$ for system size $128^2$.
Unless otherwise stated simulations were performed on a square 
lattice.
For models in the BTW class we see data 
collapse on a curve with $\gamma_{sa}=1.06$.
For the random relaxation models $\gamma_{sa}=1.23 \pm 0.01$.}
\label{figuni}
\end{figure}

In order to establish that the classification introduced above is 
a classification into universality classes we provide evidence that 
some details of the models are irrelevant 
(Fig.~\ref{figuni}).
Simulation results of the BTW model on the triangular lattice and 
square lattice were compared~\cite{manna,diaz}.
No significant difference was reported.
We define $N$ as the number of states of the $E({\bf i})$ in stable 
configurations of discrete models.
When the components of the relaxation vector are all 1's, $N$ also 
equals the number of neighbors.
In sandpile models the question of the lattice dependence or 
interaction range dependence of the exponents is actually a 
question of the dependence on $N$.
We observed a crossover effect when increasing $N$.
The scaling obtained for the BTW model on a square lattice ($N$=4) 
is shifted to larger avalanches when $N$ is increased.
Similar cross-over was observed in the other universality classes.
Note that the requirement that  ${\bf J}[\Delta E]=0$ does not 
imply isotropy.
This is the reason the universality class was called 
{\it non-directed}, rather than isotropic.
As an example, a model with a toppling vector (1,2,1,2) fulfills 
this requirement, and simulations show that it belongs to the 
universality class of
non-directed models.

Continuous models were simulated as well.
There are two types of realizations of continuous models.
In one, the variables are turned into continuous variables, and 
when the amount of sand added is not a multiple of the amount 
distributed on relaxation (or is a random variable taking such 
values) then the height profile is turned into a continuous 
distribution.
The other is the Zhang realization, where on relaxation the 
dynamic variable is decreased to zero and sand distributed equally 
among the nearest-neighbors~\cite{zhang}.
Both types seem to be in the same universality class~\cite{diaz}.
This is indicated by our simulations as well.

There is a number of possible realizations of a two-state model.
The neighbors to which sand is distributed can be chosen as 
distinct (no neighbor chosen twice) or not.
In Manna's two-state model~\cite{twostate} the variable is 
decreased to zero on relaxation, with sand distributed randomly 
among the nearest neighbors.
In this case the relaxation process depends on the variable 
value.
Continuous variants of the model may also be defined.
We have simulated realizations of such models and all were in the 
same universality class (Fig.~\ref{figuni}).
Simulations of two-state models were performed with annealed 
randomness only~\cite{randomness}.

On the basis of the wave structure of avalanches in the BTW 
model~\cite{waves}, it can be shown that avalanches have a 
``shell'' structure, i.e. the sites which relaxed at least $n+1$ 
times form a connected cluster with no holes which is contained in
 the cluster of sites which relaxed at least $n$ times 
(Fig.~\ref{figavalanche}(a)).
Avalanches in random relaxation models do not share this property, 
and their structure is more irregular.
A typical avalanche in a two-state model is shown in 
Fig.~\ref{figavalanche}(b).
These geometrical differences reflect in the fractal dimension of 
the boundary, which is greater for the two-state model.

The distinction between the universality classes of non-directed 
models and models which are non-directed only on average holds in 
three dimensions as well (Table~\ref{table3d}).
The difference is less marked because the exponents are nearing 
their mean field values.

Directed models also form a universality class.
In addition to the models studied by Dhar and Ramaswamy, 
where the relaxation 
vector is of the form (1,1,0,0) or (1,1,1,0), we simulated models 
with the relaxation vectors (1,1,1,2) and (1,1,2,2).
In the latter, multiple relaxations are possible, but it does not 
reflect in the scaling behavior.
We found the same exponent values in all these models.  
The values we obtained in simulations (Table \ref{tablegamma}) are 
in agreement with the analytical solution. 
Directed models with a random relaxation rule show cross-over. 

In summary, using extensive numerical simulations we identified 
three universality classes in sandpile models:  (a) non-directed 
models (BTW model); (b) random relaxation models which are 
non-directed only on average (Manna two-state model) and (c) 
directed models (Dhar and Ramaswamy's directed model).
These universality classes correspond to a classification according
 to the value of the current ${\bf J}[\Delta E]$ and its average.
Locally non-conservative models also show SOC and may form 
a different universality class~\cite{cascades}.

Recently Pietronero et al.~\cite{coarse-dyn} introduced a novel 
theoretical framework for calculating the exponents of sandpile 
models, in a manner which immediately reveals their universality.
Within their scheme, which is purely phenomenological, the Manna 
two-state model and the BTW model are found to be in the same 
universality class.
Its failure to distinguish between the two models indicates that 
some key ingredient is missing from their scheme.
We suspect that multiple relaxation is the missing element.
Work is now in progress to extend the procedure to include some 
form of multiple relaxation.

{\bf Acknowledgements}
We thank Z. Olami, P. Bak, E. Domany, G. Grinstein, C. Jayaprakash, 
T. Kaplan and M. Paczuski
for helpful discussions.

\begin{figure}
\caption{Simulation results for the BTW model (circles) and 
two-state model (squares).
$E[r | t]$ (average radius of gyration for given avalanche 
lifetime) vs. $t$ is displayed in (a), yielding $\gamma_{rt}$.
In (b) we show a graph of $E[s | a]$ vs. $a$ which yields 
$\gamma_{sa}$.
Their values are listed in Table~\ref{tablegamma}.
Data was binned with bin size increasing exponentially.
System size is $512^2$, with $10^7$ grains dropped. 
These results indicate that the two models belong to different 
universality classes.}
\label{figcomparison}
\end{figure}

\begin{figure}
\caption{Typical avalanche structure for the BTW model (a) and 
two-state model (b).  Grey-scale indicates the number of 
relaxations which occurred at each site during an avalanche.
White represents zero relaxations, and black represents the 
maximal no. of relaxations (10 in (a), 45 in (b)).
System size is $150^2$.
Note the shell structure in the BTW avalanche (an analytically 
provable property) vs. the irregular structure of the avalanche 
in the two-state model.
These qualitative geometrical differences translate into 
quantitative differences in exponent value, especially the fractal 
dimension of the boundary.}
\label{figavalanche}
\end{figure}

\begin{table}
\caption{$\gamma$ exponents for universality classes in two 
dimensions.
The other $\gamma$'s can be found from the scaling relations, Eq. 
[\ref{gamma}]. 
The values of these exponents were observed to be independent of 
system size.
The typical spread of data 
for different runs
of different models within the universality class
is $\pm 0.01$ about the mean.}
\label{tablegamma}
\end{table}

\begin{table}
\caption{Exponents in three dimensions for the BTW model and a 
three-state random relaxation model (non-directed on average). 
Distribution exponents are given for system size of $96^3$. }
\label{table3d}
\end{table}

\newpage

%Table I

\begin{table}
\centering
\begin{tabular}{cddd}
Exponent  &\multicolumn{3}{c}{model} \\ \cline{2-4}  
             & \multicolumn{1}{c}{BTW}   & \multicolumn{1}{c}
{two-state}  
&\multicolumn{1}{c}{Directed} \\ \hline

$1/z$\tablenote{In non-directed models $z$ is identified with
 $\gamma_{tr}$, and in directed models it is identified with 
$\gamma_{td}$.}
             &  0.76    & 0.67   & 1.00    \\ 
$\gamma_{st}$&  1.62    & 1.70   & 1.51    \\ 
$\gamma_{at}$&  1.53    & 1.35   & 1.51    \\
$\gamma_{sa}$&  1.06    & 1.23   & 1.00     \\ 
$D_f$        &  1.26    & 1.42   & 1.00     
\end{tabular}
\end{table}

\newpage

%Table II

\begin{table}
\centering
\begin{tabular}{cdd}	

Exponent     &\multicolumn{1}{c}{BTW} 
		&\multicolumn{1}{c}{3-state} \\\hline 
$\tau_s$     & 2.35 & 2.43 \\
$\tau_a$     & 2.35 & 2.46 \\
$\gamma_{rt}$
$(1/z)$	     & 0.60 & 0.54 \\ 
$\gamma_{st}$& 1.78 & 1.80 \\
$\gamma_{at}$& 1.78 & 1.72 \\
$\gamma_{sa}$& 1.00 & 1.06  

\end{tabular}
\end{table}


\begin{thebibliography}{christ-fogedby}
\bibitem{randomness}
	Simulating the two-state model with quenched randomness is
	not possible
	because the topology permits the formation of
	closed loops of relaxation.
	These loops have no outlet to the boundary and the result
	is avalanches
	which last indefinitely.
\end{thebibliography}
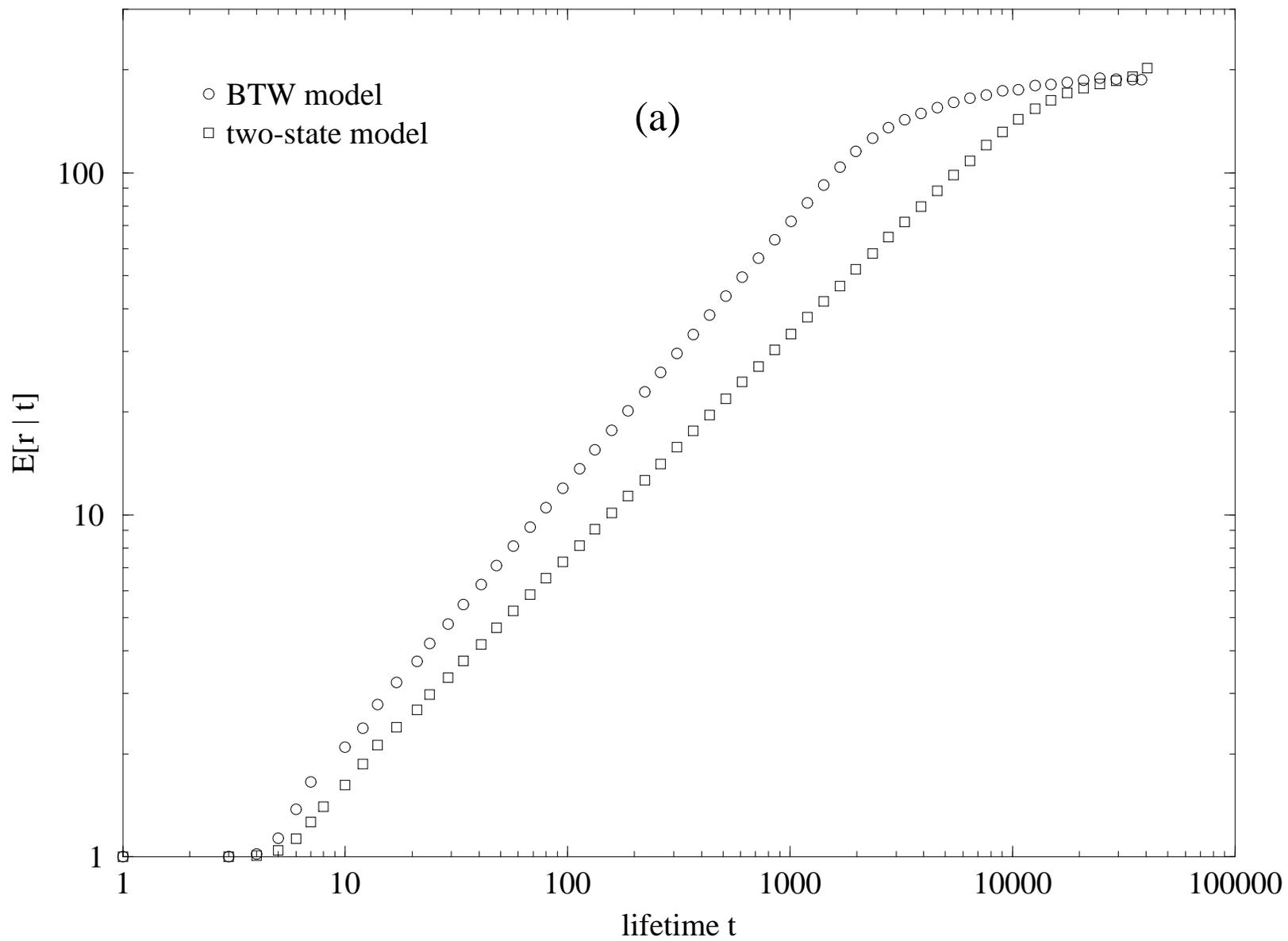
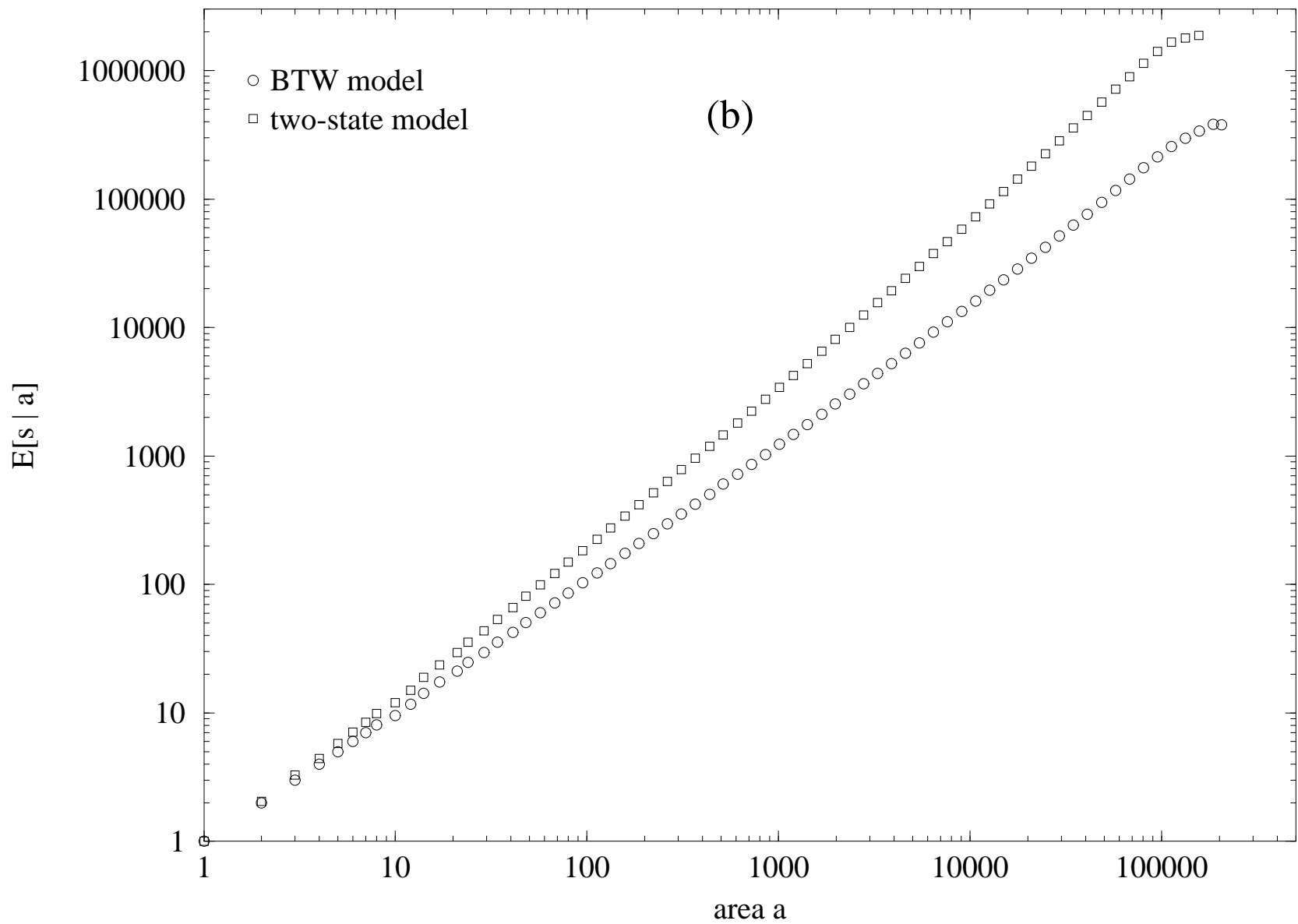
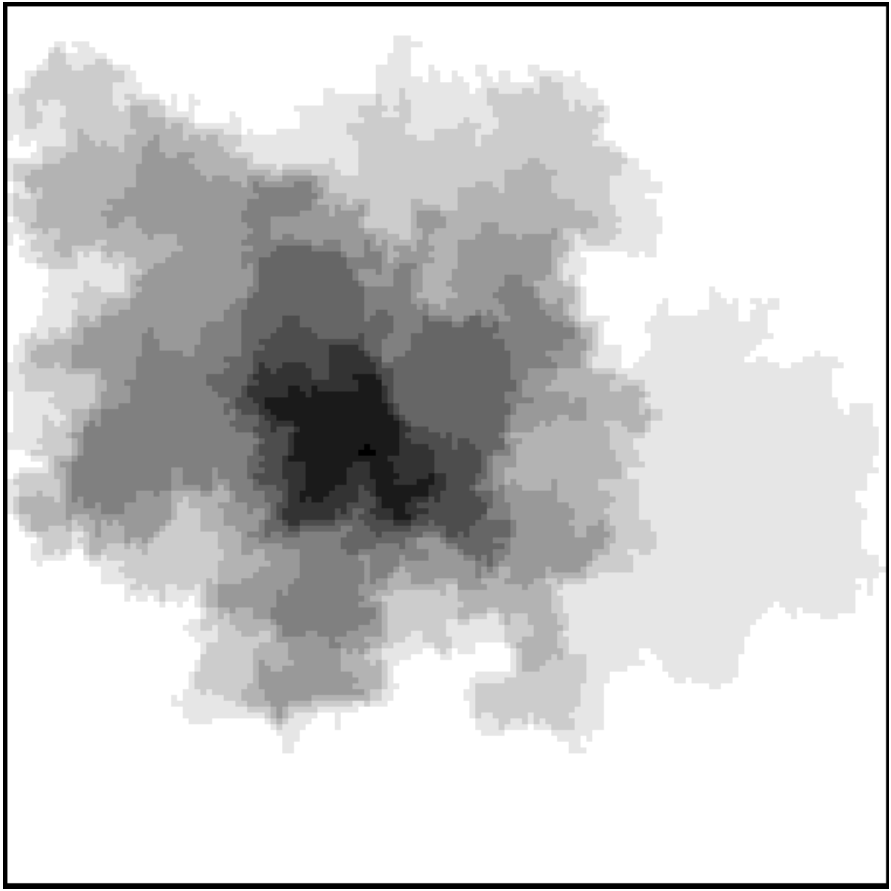
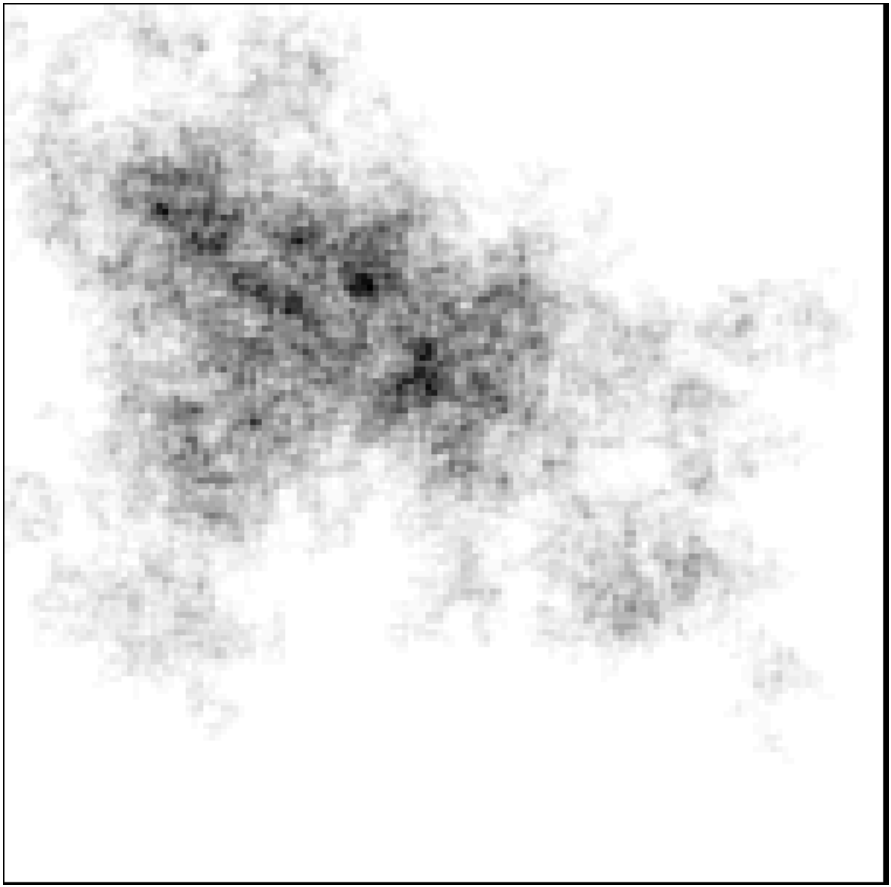
\end{document}